# Breakdown of the static picture of defect energetics in halide perovskites: the case of the Br vacancy in CsPbBr$_3$

Ayala V. Cohen,[1] David A. Egger,[2,3] Andrew M. Rappe,[4] and Leeor Kronik[1]

*[1] Department of Materials and Interfaces, Weizmann Institute of Science, Rehovoth 76100, Israel.*

*[2] Institute of Theoretical Physics, University of Regensburg, 93040 Regensburg, Germany.*

*[3] Department of Physics, Technical University of Munich, 85748 Garching, Germany*

*[4]Department of Chemistry, University of Pennsylvania, Philadelphia, PA 19104–6323, USA.*

**Abstract**

We consider the Br vacancy in CsPbBr$_3$ as a prototype for the impact of structural dynamics on defect energetics in halide perovskites (HaPs). Using first-principles molecular dynamics based on density functional theory, we find that the static picture of defect energetics breaks down; the energy of the V$_{Br}$ level is found to be intrinsically dynamic, oscillating by as much as 1 eV on the ps time scale at room temperature. These significant energy fluctuations are correlated with the distance between the neighboring Pb atoms across the vacancy and with the electrostatic potential at these Pb atomic sites. We expect this unusually strong coupling of structural dynamics and defect energetics to bear important implications for both experimental and theoretical analysis of defect characteristics in HaPs. It may also hold significant ramifications for carrier transport and defect tolerance in this class of photovoltaic materials.

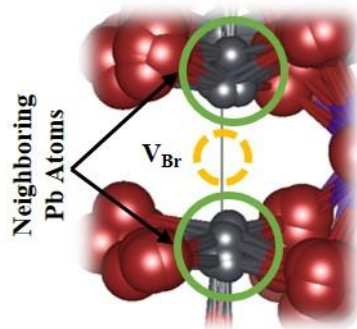

Halide perovskites (HaPs) have emerged as remarkably efficient materials for optoelectronic applications, notably for photovoltaics, but also for light emitting diodes and other devices.[1–8] One of the most intriguing features of HaPs is the very long lifetime of photoexcited carriers (up to several μs),[9–11] which is essential for key device parameters, *e.g.*, the open-circuit voltage of HaP-based photovoltaic cells.[12] These long lifetimes are particularly remarkable given that HaP compounds are typically fabricated using relatively low-energy, solution-based material syntheses.[13–16] This suggests that electronic defect states in these materials are unusually benign electrically.[10,17–21] Therefore, there are significant and on-going efforts to understand defect energetics and its relation to charge carrier scattering and recombination in HaP materials.

Recently, a large number of theoretical investigations have examined the microscopic origins and energetics of intrinsic defects in HaPs, mostly using density functional theory (DFT).[22-51] While these studies exhibit scatter in the results owing to the type of density functional used, as well as various numerical and convergence choices, they do share one key common element, which is the adoption of a static model. In other words, these studies allow the surrounding nuclei to adapt to the presence of a defect via a well-defined local minimum in the potential energy surface, tacitly assuming that dynamic nuclear effects are of secondary importance. Comparison between theory and experiment shows this assumption to be justified in a wide range of semiconductors of scientific and technological interest.[52] However, it is by now apparent that HaPs do not behave as ordinary semiconductors in terms of lattice dynamics. They are mechanically much softer than other efficient optoelectronic materials,[21,53–59] and exhibit unusual structural dynamical phenomena including large-amplitude, anharmonic nuclear fluctuations and dynamic symmetry breaking, even at room temperature (RT).[21,58,60–90] Given that such phenomena surely modulate the energy landscape both spatially and temporally, one may ask whether the assumption of a static defect is valid, and if not, what are the consequences of its breakdown.

Here, we address this question by performing first-principles molecular dynamics (MD) calculations, based on DFT, which allow us to study point defect properties while fully accounting for structural dynamics. As an illuminating test case, we focus on $CsPbBr_3$, which is known to exhibit structural dynamics similar to hybrid organic-inorganic HaPs,[68] and we examine the halide vacancy, which has been suggested theoretically to be abundant.[36,91] We find significant temporal variation in defect energies which are strongly correlated with nuclear dynamics, thereby precluding the assignment of a unique defect energy level. This suggests a breakdown of the static picture, with possible repercussions for material and device behavior.

All DFT-based MD calculations were performed using the Perdew-Burke-Ernzerhof (PBE)[92] form of the generalized-gradient approximation, augmented by dispersion terms calculated within the Tkatchenko-Scheffler (TS) scheme[93] using an iterative Hirshfeld partitioning of the self-consistently computed charge density.[94] This approach has previously been found to be highly useful for an accurate description of the structure and energetics of ionic compounds that also exhibit dispersive binding.[94,95] We note that the Heyd-Scuseria-Ernzerhof (HSE)[96] short-range hybrid functional, together with the inclusion of spin-orbit coupling (SOC) effects, has previously been shown to be more accurate in predicting defect energetics in HaPs.[31] However, a comprehensive MD analysis of the system studied here based on HSE+SOC is at present prohibitive computationally. Here, our strongest interest is not in the absolute defect level position, but rather in its energy fluctuations triggered by thermal nuclear motions. As shown in the supporting information (SI), Section 1, all energy fluctuation trends reported below are fully reproduced by HSE+SOC calculations performed for selected points in the MD run, thus validating the use of PBE level calculations in this case. All calculations are performed using the Vienna ab initio simulation package (VASP),[97] a plane-wave basis code in which ionic cores are described by the projected augmented wave (PAW)[98] method. A plane-wave cutoff of 300 eV and a $10^{-4}$ eV/supercell convergence criterion for the total energy were used in all HaP calculations.

All first-principles MD simulations of $CsPbBr_3$ in its RT orthorhombic phase were based on previously used cell parameters ($a$=8.18Å, $b$=11.64Å, $c$=8.13Å),[68] obtained from a PBE+TS optimization of the ionic and cell degrees of freedom. Whereas PBE alone overestimates cell parameters, the inclusion of dispersive TS terms greatly improves the agreement between experimental and theoretical lattice parameters,[99–102] in our case to within ≈0.8% of the reported experimental values.[103] A 2×2×2 supercell (containing 160 atoms) was employed and a single Br vacancy ($V_{Br}$) was introduced. Owing to the size of this supercell, only the Γ-point of the Brillouin zone was sampled and was found to yield converged differences in defect-state eigenvalues (see below). The DFT-MD calculations were performed using a Nosé-Hoover thermostat[104,105] with the temperature set to 298 K and a Verlet algorithm to integrate the nuclear equations of motion, using a timestep of 10 fs, previously found to be sufficient for MD of CsPbBr$_3$.[68] To ascertain the suitability of this time step, we repeated the MD run of the defect-containing supercell with a significantly smaller time step of 2 fs. The results of this comparison, given in Section 2 of the SI, establish that the two time steps result in similar statistics. CsPbBr$_3$ was thermally equilibrated, followed by 180 ps of RT dynamics. For comparison, similar MD calculations have been performed for the well-known As$_{Ga}$ anti-site defect in GaAs, using a 3×3×3 GaAs supercell, which provides a similar defect-defect image distance to the HaP supercell, with all other parameters being the same except for a slightly smaller cutoff energy of 280 eV. Additional tests of the effects of denser $k$-grids, of spin-polarization, and of supercell size (defect density and image interaction), are presented in Sections 3 and 4 of the SI.

The standard, static approach to computing defect level energetics is to consider the cross-over point in the formation energy of the relaxed defect at different charge states.[52] Because our MD data inherently and purposefully produce unrelaxed geometries, we focus instead on the instantaneous eigenvalue of the defect state as a proxy to its instantaneous charge transition level. We are well aware that Kohn-Sham eigenvalues are not charge transition levels, but we also emphasize that the two are qualitatively and often semi-quantitatively related,[106] an issue elaborated below. For simplicity and clarity, further use of the term "defect level" refers to the Kohn-Sham eigenvalue related to the defect state.

Fig. 1a shows the temporal evolution of three key eigenvalues in the CsPbBr$_3$ supercell containing a neutral V$_{Br}$ defect: the conduction band minimum (CBM), the valence band maximum (VBM), and the defect-related eigenvalue, over the last 30 ps of the equilibrated MD simulation. The energy variations of the VBM and CBM energies over the entire 180 ps of the MD simulation are found to be ≈0.3-0.4 eV, a result that is consistent with prior studies.[62,64,107] Also consistent with prior studies is the fact that the fluctuations of these bulk properties somewhat diminish further with supercell size, but remain finite[108] (See SI, section 4, for more details). In stark contrast, the temporal fluctuations of the V$_{Br}$–related defect energy level, and more importantly, the fluctuations in the distance of this level from the CBM, are much larger, and are much more weakly affected by an increase in supercell size due to the localization of the defect state (an issue elaborated below). The defect energy level fluctuations can be as large as ≈1.3 eV (see Table 1 for detailed statistics), often on a time scale < 1 ps. As a consequence, the defect level spans a surprisingly wide energy range, from almost mid-gap to almost the CBM. This wide range is reminiscent of the one found computationally in MAPbI$_3$, upon taking into account different MA orientations around an I vacancy under excited-state conditions.[109] For comparison, Fig. 1b shows the results of analogous calculations of the same three eigenvalues for a GaAs supercell containing a neutral As$_{Ga}$ defect. The fluctuations in the band edge energies are smaller (≈0.1-0.2 eV) than in CsPbBr$_3$, but more importantly, the fluctuations in defect energetics are significantly smaller, only ≈0.3 eV.

It is important to note that the PBE calculations of lead-based HaPs benefit from a fortuitous cancellation of errors; the combination of the PBE underestimation of the gap and an overestimation of the gap due to neglect of SOC combine to produce a band gap that is close to the experimental one.[110] In GaAs, the effect of SOC on the gap is much less pronounced, and therefore the PBE band gap (0.53 eV) is much smaller than either the experimental GaAs band gap or the HaP band gap. To check whether this affects our conclusions, we performed additional HSE+SOC calculations for GaAs, using selected geometries from the MD run (see SI, section 1, for more details). The band gap of GaAs then increases to ≈1.3 eV, close to the experimental gap value, causing the defect-level energy fluctuation to increase by a factor of two. This is larger than what would have been naively expected, but still much smaller than the ≈1.3 eV range found for the softer and more anharmonic HaP.

Importantly, the defect-level fluctuations in CsPbBr$_3$ and GaAs do not appear on the same time scale. This is established in Fig. 2, where the fluctuations are averaged over a temporal "window". Upon averaging the fluctuations over a temporal window of 100 fs, a time scale that is assumed to be relevant for carrier scattering or trapping events, the range of defect fluctuations in GaAs reduces to a mere ≈0.1 eV, whereas defect fluctuations in CsPbBr$_3$ still span over ≈1.2 eV. Averaging over a significant temporal range of 1 ps (also shown in Fig. 2) causes defect fluctuations in GaAs to be essentially negligible, whereas in CsPbBr$_3$ they still span a significant ≈0.6 eV; they remain non-negligible even upon averaging over a window of 5 ps. This rapid convergence towards one unique value explains why in GaAs the static picture can still be used successfully to address defect properties, whereas this picture breaks down in CsPbBr$_3$.

In order to ascertain that these results are not a spurious consequence of considering the defect eigenvalues instead of the traditional charge transition levels, we compute the total energies of all geometries along a 15 ps section of the MD run, while adding a positive charge to the defect. The total-energy differences between the neutral defect-containing supercell and the positively-charged one produce the vertical transition levels along the trajectory, i.e., transition energies in the absence of atomic relaxation upon charge capture or release. A comparison between the eigenvalues and vertical transition levels is presented in Fig. 3. As expected theoretically, the two quantities are not entirely the same, but trends in the fluctuation of both properties are highly correlated, thus supporting the choice to focus on the defect eigenvalues as a proxy. We note that the comparison to the vertical transition level, as opposed to the thermodynamic (relaxed) one, is indeed the relevant one, as the defect eigenvalues do not contain information as to structural relaxation upon charge capture or release. For a detailed analysis, see Section 5 in the SI.

In order to understand the microscopic origins of the unusual fluctuations in the energetics of $V_{Br}$ in $CsPbBr_3$, we calculate the correlations of these fluctuations with fluctuations in three other key quantities along the MD trajectory: the CBM, the distance between the two Pb atoms adjacent to the Br vacancy, and the electrostatic energy at either of these two lattice sites, gauged from their calculated $1s$ core level energy. These results are given in Fig. 4. Clearly, the defect-level fluctuations are found to be reasonably correlated with those of the CBM (correlation coefficient: 0.46). However, they are uncorrelated with fluctuations in the VBM (correlation coefficient: 0.01, see SI, Section 6). This observation is rationalized by the similar chemical nature of the $n$-type defect state and the conduction band: the Br vacancy results in dangling $p$-orbitals for the two Pb atoms surrounding it, and the conduction band also comprises Pb $p$-orbitals. Importantly, the fluctuations in the defect level are also highly correlated with fluctuations in both the distance between the adjacent Pb atoms (correlation coefficient: 0.72) and the electrostatic potential at these Pb sites (correlation coefficient: 0.54). Significantly weaker correlation (*i.e.*, correlation coefficients smaller than 0.3) was found with fluctuations in other neighboring atoms or the electrostatic potential at their sites (see SI, section 6).

The strong correlation of the fluctuations in defect energetics with those in the position and potential of the neighboring Pb atoms strongly motivates study of the electronic structure changes that accompany these Pb motions. To gain further insight into this issue, we examine five geometries of the $V_{Br}$ defect, selected from the MD trajectory, to represent the entire range of defect energy fluctuations. Generally, we find that a smaller Pb-Pb distance is associated with a deeper trap, whereas a larger Pb-Pb distance is associated with a shallower one. In Fig. 5a, the geometries of the five snapshots are overlaid, showing that strong Pb motion is indeed a dominant geometrical feature in the vicinity of the defect. Fig. 5b then shows the charge density associated with the defect state for the five snapshots. Clearly, deeper defect energies (smaller Pb-Pb distances) are associated with localization of the defect state, whereas shallower defect states (larger Pb-Pb distances) are more delocalized. This interpretation is further supported by MD runs of the supercell containing a positively charged Br vacancy (see SI, Section 7, for details). While the overall findings are similar to the case of the neutral species, and the energy fluctuations of the defect level are still very large, the range is only about half that of the neutral defect. This makes sense if one considers that the removal of an electron creates more repulsion between the positively charged Pb atoms and limits their range of motion to areas farther away from the vacancy. This reduced range of motion then translates to reduced fluctuations in the energy of the defect state.

These results, for both charge states of the defect, leave us with one very important insight – defect energetics in HaPs are dynamic. Therefore, pinning down a particular unique static value, even if highly precise, is not sufficiently informative. The strong spatial and temporal modulation of the energetics in HaPs is *dictated by the local and instantaneous arrangement of atoms rather than by one global static geometry,* causing the defect energetics to span a significant portion of the forbidden band gap on the ps time scale. This effect has been demonstrated here for the case of the $V_{Br}$ defect in $CsPbBr_3$, which is an abundant defect in HaPs. The situation may well be similar for other types of defects in view of the soft HaP lattice. Of course, each defect should be studied individually, but the large defect-level fluctuations found in our example are fundamentally incompatible with the static picture and suffice to establish that it cannot be taken for granted *a priori*.

Our results may have important implications for future studies on HaPs as well as potential ramifications for devices. This is because the unusually long time scale of the defect energy fluctuations, as well as the fact that it changes strongly but remains large when considering longer times (see Fig. 2 above), implies a frequency dependence in the response to measurements probing the defect characteristics of HaPs. This poses significant challenges also to experimental defect characterization of HaPs, where fluctuations over long enough time scales may affect measured thermal or optical properties. Finally, we hypothesize that defect fluctuations could facilitate a mechanism for defect tolerance: While the average defect energy position may be deep in the gap and thus likely to capture charge carriers and be harmful for solar cell operation, the DFT-MD results imply that the structural fluctuations of the environment can also lead to frequent visits to regions of shallow defect energies. These are less likely to trap carriers and could release captured ones, which could potentially mitigate deleterious trapping effects of the defect. This suggested mechanism bears some resemblance to the "electron elevator", suggested by Lim *et al.*[111] who describe the movement of a gap state energy level across the forbidden gap by moving ions in irradiated Si. We believe that the results presented in this work call for further study of this possibility.

In conclusion, we have considered the case of the Br vacancy in $CsPbBr_3$ as a prototype for examining the impact of pertinent nuclear fluctuations at room temperature on defect energetics in HaPs. Our DFT-MD results show that the static picture of defects breaks down in $CsPbBr_3$; the energy of the $V_{Br}$ level is dynamic. It oscillates widely and over long time scales within the forbidden band gap, which is unusual compared to conventional semiconductors such as GaAs. The energy fluctuations of the defect state were found to be correlated with the distance of neighboring Pb atoms as well as with the surrounding electrostatic potential. Therefore, the massive energy fluctuations of the $V_{Br}$ defect in $CsPbBr_3$ are dictated by the large-amplitude nuclear displacements of the HaP lattice. These findings bear important implications for future experimental and theoretical studies on defect characteristics of HaPs, as well as their consequences in devices, and may specifically assist defect tolerance in these materials.

**Acknowledgements**

We thank Professors Chris van de Walle (UCSB), Aron Walsh (Imperial College), Feliciano Giustino (Oxford), and Alessandro Troisi (U Liverpool), for illuminating discussions and constructive criticism. AVC and LK acknowledge support by the Minerva Foundation. DAE acknowledges funding provided by the Alexander von Humboldt Foundation in the framework of the Sofja Kovalevskaja Award endowed by the German Federal Ministry of Education and Research is acknowledged. AMR acknowledges support from the US National Science Foundation, under grant DMR-1719353. L.K. is the incumbent of the Aryeh and Mintzi Katzman Professorial Chair.

**References**

1.  Brenner, T. M., Egger, D. A., Kronik, L., Hodes, G. & Cahen, D. Hybrid organic - Inorganic perovskites: Low-cost semiconductors with intriguing charge-transport properties. *Nat. Rev. Mater.* **1,** 15007 (2016).

2.  Li, W. *et al.* Chemically diverse and multifunctional hybrid organic-inorganic perovskites. *Nat. Rev. Mater.* **2,** 16099 (2017).

3.  Ono, L. K., Juarez-Perez, E. J. & Qi, Y. Progress on Perovskite Materials and Solar Cells with Mixed Cations and Halide Anions. *ACS Appl. Mater. Interfaces* **9,** 30197–30246 (2017).

4.  Ahmadi, M., Wu, T. & Hu, B. A Review on Organic–Inorganic Halide Perovskite Photodetectors: Device Engineering and Fundamental Physics. *Adv. Mater.* **29,** 1605242 (2017).

5.  Correa-Baena, J. P. *et al.* Promises and challenges of perovskite solar cells. *Science* **358,** 739–744 (2017).

6.  Kovalenko, M. V., Protesescu, L. & Bodnarchuk, M. I. Properties and potential optoelectronic applications of lead halide perovskite nanocrystals. *Science* **358,** 745–750 (2017).

7.  Van Le, Q., Jang, H. W. & Kim, S. Y. Recent Advances toward High-Efficiency Halide Perovskite Light-Emitting Diodes: Review and Perspective. *Small Methods* 1700419 (2018).

8.  Leijtens, T., Bush, K. A., Prasanna, R. & McGehee, M. D. Opportunities and challenges for tandem solar cells using metal halide perovskite semiconductors. *Nat. Energy* **3,** 1–11 (2018).

9.  Wehrenfennig, C., Eperon, G. E., Johnston, M. B., Snaith, H. J. & Herz, L. M. High charge carrier mobilities and lifetimes in organolead trihalide perovskites. *Adv. Mater.* **26,** 1584–1589 (2014).

10. Shi, D. *et al.* Low trap-state density and long carrier diffusion in organolead trihalide perovskite single crystals. *Science* **347,** 519–522 (2015).

11. Stranks, S. D. *et al.* Electron-Hole Diffusion Lengths Exceeding 1 Micrometer

in an Organometal Trihalide Perovskite Absorber. *Science* **342,** 341–344 (2018).

12. Tress, W. *et al.* Predicting the Open-Circuit Voltage of $CH_3NH_3PbI_3$ Perovskite Solar Cells Using Electroluminescence and Photovoltaic Quantum Efficiency Spectra: the Role of Radiative and Non-Radiative Recombination. *Adv. Energy Mater.* **5,** 1400812 (2015).

13. Zhao, Y. & Zhu, K. Solution chemistry engineering toward high-efficiency perovskite solar cells. *J. Phys. Chem. Lett.* **5,** 4175–4186 (2014).

14. Stranks, S. D., Nayak, P. K., Zhang, W., Stergiopoulos, T. & Snaith, H. J. Formation of thin films of organic-inorganic perovskites for high-efficiency solar cells. *Angew. Chemie - Int. Ed.* **54,** 3240–3248 (2015).

15. Yan, K. *et al.* Hybrid Halide Perovskite Solar Cell Precursors: Colloidal Chemistry and Coordination Engineering behind Device Processing for High Efficiency. *J. Am. Chem. Soc.* **137,** 4460–4468 (2015).

16. Nie, W. *et al.* High-efficiency solution-processed perovskite solar cells with millimeter-scale grains. *Science* **347,** 522–525 (2015).

17. Xing, G. *et al.* Low-temperature solution-processed wavelength-tunable perovskites for lasing. *Nat. Mater.* **13,** 476–480 (2014).

18. Brandt, R. E., Stevanović, V., Ginley, D. S. & Buonassisi, T. Identifying defect-tolerant semiconductors with high minority-carrier lifetimes: Beyond hybrid lead halide perovskites. *MRS Commun.* **5,** 265–275 (2015).

19. Rosenberg, J. W., Legodi, M. J., Rakita, Y., Cahen, D. & Diale, M. Laplace current deep level transient spectroscopy measurements of defect states in methylammonium lead bromide single crystals. *J. Appl. Phys.* **122,** 145701 (2017).

20. Kirchartz, T., Markvart, T., Rau, U. & Egger, D. A. Impact of Small Phonon Energies on the Charge-Carrier Lifetimes in Metal-Halide Perovskites. *J. Phys. Chem. Lett.* **9,** 939–946 (2018).

21. Egger, D. A. *et al.* What Remains Unexplained about the Properties of Halide Perovskites? *Adv. Mater.* **30,** 1–11 (2018).

22. Chung, I. *et al.* $CsSnI_3$: Semiconductor or metal? High electrical conductivity and strong near-infrared photoluminescence from a single material. High hole mobility and phase-transitions. *J. Am. Chem. Soc.* **134,** 8579–8587 (2012).

23. Buin, A. *et al.* Materials processing routes to trap-free halide perovskites. *Nano*

*Lett.* **14,** 6281–6286 (2014).

24. Du, M. H. Efficient carrier transport in halide perovskites: Theoretical perspectives. *J. Mater. Chem. A* **2,** 9091–9098 (2014).

25. Yin, W. J., Shi, T. & Yan, Y. Unusual defect physics in $CH_3NH_3PbI_3$ perovskite solar cell absorber. *Appl. Phys. Lett.* **104,** 063903/1--063903/4 (2014).

26. Agiorgousis, M. L., Sun, Y. Y., Zeng, H. & Zhang, S. Strong covalency-induced recombination centers in perovskite solar cell material $CH_3NH_3PbI_3$. *J. Am. Chem. Soc.* **136,** 14570–14575 (2014).

27. Xu, P., Chen, S., Xiang, H. J., Gong, X. G. & Wei, S. H. Influence of defects and synthesis conditions on the photovoltaic performance of perovskite semiconductor $CsSnI_3$. *Chem. Mater.* **26,** 6068–6072 (2014).

28. Shi, H. & Du, M. H. Shallow halogen vacancies in halide optoelectronic materials. *Phys. Rev. B - Condens. Matter Mater. Phys.* **90,** 1–6 (2014).

29. Walsh, A., Scanlon, D. O., Chen, S., Gong, X. G. & Wei, S. H. Self-regulation mechanism for charged point defects in hybrid halide perovskites. *Angew. Chemie - Int. Ed.* **54,** 1791–1794 (2015).

30. Shi, T., Yin, W. J., Hong, F., Zhu, K. & Yan, Y. Unipolar self-doping behavior in perovskite $CH_3NH_3PbBr_3$. *Appl. Phys. Lett.* **106,** 3–8 (2015).

31. Du, M. H. Density functional calculations of native defects in $CH_3NH_3PbI_3$: Effects of spin - Orbit coupling and self-interaction error. *J. Phys. Chem. Lett.* **6,** 1461–1466 (2015).

32. Xiao, Z., Zhou, Y., Hosono, H. & Kamiya, T. Intrinsic defects in a photovoltaic perovskite variant $Cs_2SnI_6$. *Phys. Chem. Chem. Phys.* **17,** 18900–18903 (2015).

33. Yang, J. H., Yin, W. J., Park, J. S. & Wei, S. H. Self-regulation of charged defect compensation and formation energy pinning in semiconductors. *Sci. Rep.* **5,** 16977 (2015).

34. Kim, J., Chung, C. H. & Hong, K. H. Understanding of the formation of shallow level defects from the intrinsic defects of lead tri-halide perovskites. *Phys. Chem. Chem. Phys.* **18,** 27143–27147 (2016).

35. Mosconi, E., Meggiolaro, D., Snaith, H. J., Stranks, S. D. & De Angelis, F. Light-induced annihilation of Frenkel defects in organo-lead halide perovskites. *Energy Environ. Sci.* **9,** 3180–3187 (2016).

36. Kang, J. & Wang, L. W. High Defect Tolerance in Lead Halide Perovskite

CsPbBr$_3$. *J. Phys. Chem. Lett.* **8,** 489–493 (2017).

37.    Whalley, L. D., Crespo-Otero, R. & Walsh, A. H-Center and V-Center Defects in Hybrid Halide Perovskites. *ACS Energy Lett.* **2,** 2713–2714 (2017).

38.    Li, W., Liu, J., Bai, F. Q., Zhang, H. X. & Prezhdo, O. V. Hole Trapping by Iodine Interstitial Defects Decreases Free Carrier Losses in Perovskite Solar Cells: A Time-Domain Ab Initio Study. *ACS Energy Lett.* **2,** 1270–1278 (2017).

39.    Meggiolaro, D. *et al.* Iodine chemistry determines the defect tolerance of lead-halide perovskites. *Energy Environ. Sci.* **11,** 702–713 (2018).

40.    Walukiewicz, W. *et al.* Bistable Amphoteric Native Defect Model of Perovskite Photovoltaics. *J. Phys. Chem. Lett.* **9,** 3878–3885 (2018).

41.    Saidaminov, M. I. *et al.* Suppression of atomic vacancies via incorporation of isovalent small ions to increase the stability of halide perovskite solar cells in ambient air. *Nat. Energy* **3,** 648–654 (2018).

42.    Meggiolaro, D., Mosconi, E. & De Angelis, F. Modeling the Interaction of Molecular Iodine with MAPbI$_3$: A Probe of Lead-Halide Perovskites Defect Chemistry. *ACS Energy Lett.* **3,** 447–451 (2018).

43.    Li, T., Zhao, X., Yang, D., Du, M. H. & Zhang, L. Intrinsic Defect Properties in Halide Double Perovskites for Optoelectronic Applications. *Phys. Rev. Appl.* **10,** 41001 (2018).

44.    Wong, A. B. *et al.* Strongly Quantum Confined Colloidal Cesium Tin Iodide Perovskite Nanoplates: Lessons for Reducing Defect Density and Improving Stability. *Nano Lett.* **18,** 2060–2066 (2018).

45.    Wiktor, J., Ambrosio, F. & Pasquarello, A. Mechanism suppressing charge recombination at iodine defects in CH$_3$NH$_3$PbI$_3$ by polaron formation. *J. Mater. Chem. A* **6,** 16863–16867 (2018).

46.    Liu, N. & Yam, C. Y. First-principles study of intrinsic defects in formamidinium lead triiodide perovskite solar cell absorbers. *Phys. Chem. Chem. Phys.* **20,** 6800–6804 (2018).

47.    Brenes, R., Eames, C., Bulović, V., Islam, M. S. & Stranks, S. D. The Impact of Atmosphere on the Local Luminescence Properties of Metal Halide Perovskite Grains. *Adv. Mater.* **30,** 1706208 (2018).

48.    Park, J. S., Kim, S., Xie, Z. & Walsh, A. Point defect engineering in thin-film solar cells. *Nat. Rev. Mater.* **3,** 194–210 (2018).

49.  Yavari, M. *et al.* How far does the defect tolerance of lead-halide perovskites range? The example of Bi impurities introducing efficient recombination centers. *J. Mater. Chem. A* (2019).

50.  Meggiolaro, D., Mosconi, E. & De Angelis, F. Formation of Surface Defects Dominates Ion Migration in Lead-Halide Perovskites. *ACS Energy Lett.* **4,** 779–785 (2019).

51.  Li, J. L., Yang, J., Wu, T. & Wei, S. H. Formation of DY center as n-type limiting defects in octahedral semiconductors: The case of Bi-doped hybrid halide perovskites. *J. Mater. Chem. C* **7,** 4230–4234 (2019).

52.  Freysoldt, C. *et al.* First-principles calculations for point defects in solids. *Rev. Mod. Phys.* **86,** 253–305 (2014).

53.  Chang, Y. H., Park, C. H. & Matsuishi, K. First-Principles Study of the Structural and the Electronic Properties of the Lead-Halide-Based Inorganic-Organic Perovskites ($CH_3NH_3$)$PbX_3$ and $CsPbX_3$ (X = Cl, Br, I). *J. Korean Phys. Soc.* **44,** 889–893 (2004).

54.  Feng, J. Mechanical properties of hybrid organic-inorganic $CH_3NH_3BX_3$ (B = Sn, Pb; X = Br, I) perovskites for solar cell absorbers. *APL Mater.* **2,** 81801 (2014).

55.  Rakita, Y., Cohen, S. R., Kedem, N. K., Hodes, G. & Cahen, D. Mechanical properties of $APbX_3$ (A = Cs or $CH_3NH_3$; X = I or Br) perovskite single crystals. *MRS Commun.* **5,** 623–629 (2015).

56.  Sun, S., Fang, Y., Kieslich, G., White, T. J. & Cheetham, A. K. Mechanical properties of organic-inorganic halide perovskites, $CH_3NH_3PbX_3$ (X = I, Br and Cl), by nanoindentation. *J. Mater. Chem. A* **3,** 18450–18455 (2015).

57.  Lee, J.-H., Deng, Z., Bristowe, N. C., Bristowe, P. D. & Cheetham, A. K. The competition between mechanical stability and charge carrier mobility in MA-based hybrid perovskites: insight from DFT. *J. Mater. Chem. C* (2018).

58.  Kabakova, I. V. *et al.* The effect of ionic composition on acoustic phonon speeds in hybrid perovskites from Brillouin spectroscopy and density functional theory. *J. Mater. Chem. C* **6,** 3861–3868 (2018).

59.  Ferreira, A. C. *et al.* Elastic Softness of Hybrid Lead Halide Perovskites. *Phys. Rev. Lett.* **121,** 085502 (2018).

60.  Poglitsch, A. & Weber, D. Dynamic disorder in methylammoniumtrihalogenoplumbates (II) observed by millimeter-wave

spectroscopy. *J. Chem. Phys.* **87,** 6373–6378 (1987).

61.     Worhatch, R. J., Kim, H. J., Swainson, I. P., Yonkeu, A. L. & Billinge, S. J. L. Study of local structure in selected organic-inorganic perovskites in the Pm3̄m phase. *Chem. Mater.* **20,** 1272–1277 (2008).

62.     Carignano, M. A., Kachmar, A. & Hutter, J. Thermal effects on $CH_3NH_3PbI_3$ perovskite from Ab initio molecular dynamics simulations. *J. Phys. Chem. C* **119,** 8991–8997 (2015).

63.     Motta, C., El-Mellouhi, F. & Sanvito, S. Charge carrier mobility in hybrid halide perovskites. *Sci. Rep.* **5,** 12746 (2015).

64.     Motta, C. *et al.* Revealing the role of organic cations in hybrid halide perovskite $CH_3NH_3PbI_3$. *Nat. Commun.* **6,** 7026 (2015).

65.     Egger, D. A., Rappe, A. M. & Kronik, L. Hybrid Organic-Inorganic Perovskites on the Move. *Acc. Chem. Res.* **49,** 573–581 (2016).

66.     Beecher, A. N. *et al.* Direct Observation of Dynamic Symmetry Breaking above Room Temperature in Methylammonium Lead Iodide Perovskite. *ACS Energy Lett.* **1,** 880–887 (2016).

67.     Sendner, M. *et al.* Optical phonons in methylammonium lead halide perovskites and implications for charge transport. *Mater. Horizons* **3,** 613–620 (2016).

68.     Yaffe, O. *et al.* Local Polar Fluctuations in Lead Halide Perovskite Crystals. *Phys. Rev. Lett.* **118,** 136001 (2017).

69.     Wu, X. *et al.* Light-induced picosecond rotational disordering of the inorganic sublattice in hybrid perovskites. *Sci. Adv.* **3,** 1602388 (2017).

70.     Miyata, K. *et al.* Large polarons in lead halide perovskites. *Sci. Adv.* **3,** 1701217 (2017).

71.     Guo, Y. *et al.* Interplay between organic cations and inorganic framework and incommensurability in hybrid lead-halide perovskite $CH_3NH_3PbBr_3$. *Phys. Rev. Mater.* **1,** 042401 (2017).

72.     Yang, R. X., Skelton, J. M., Da Silva, E. L., Frost, J. M. & Walsh, A. Spontaneous octahedral tilting in the cubic inorganic cesium halide perovskites $CsSnX_3$ and $CsPbX_3$ (X = F, Cl, Br, I). *J. Phys. Chem. Lett.* **8,** 4720–4726 (2017).

73.     Fabini, D. H. *et al.* Universal Dynamics of Molecular Reorientation in Hybrid Lead Iodide Perovskites. *J. Am. Chem. Soc.* **139,** 16875–16884 (2017).

74. Uratani, H. & Yamashita, K. Inorganic Lattice Fluctuation Induces Charge Separation in Lead Iodide Perovskites: Theoretical Insights. *J. Phys. Chem. C* **121,** 26648–26654 (2017).

75. Capitani, F. *et al.* Locking of Methylammonium by Pressure-Enhanced H-Bonding in (CH₃NH₃)PbBr₃ Hybrid Perovskite. *J. Phys. Chem. C* **121,** 28125–28131 (2017).

76. Niesner, D. *et al.* Structural fluctuations cause spin-split states in tetragonal (CH₃NH₃)PbI₃ as evidenced by the circular photogalvanic effect. *Proc. Natl. Acad. Sci.* **115,** 9509–9514 (2018).

77. Schueller, E. C. *et al.* Crystal Structure Evolution and Notable Thermal Expansion in Hybrid Perovskites Formamidinium Tin Iodide and Formamidinium Lead Bromide. *Inorg. Chem.* **57,** 695–701 (2018).

78. Maughan, A. E. *et al.* Anharmonicity and Octahedral Tilting in Hybrid Vacancy-Ordered Double Perovskites. *Chem. Mater.* **30,** 472–483 (2018).

79. Bechtel, J. S. & Van der Ven, A. Octahedral tilting instabilities in inorganic halide perovskites. *Phys. Rev. Mater.* **2,** 025401 (2018).

80. Marronnier, A. *et al.* Anharmonicity and Disorder in the Black Phases of Cesium Lead Iodide Used for Stable Inorganic Perovskite Solar Cells. *ACS Nano* **12,** 3477–3486 (2018).

81. McKechnie, S. *et al.* Dynamic symmetry breaking and spin splitting in metal halide perovskites. *Phys. Rev. B* **98,** 085108 (2018).

82. Zhang, X., Shen, J. X., Wang, W. & Van De Walle, C. G. First-Principles Analysis of Radiative Recombination in Lead-Halide Perovskites. *ACS Energy Lett.* **3,** 2329–2334 (2018).

83. Thouin, F. *et al.* Phonon coherences reveal the polaronic character of excitons in two-dimensional lead halide perovskites. *Nat. Mater.* **18,** 349–356 (2019).

84. Guo, Y. *et al.* Dynamic emission Stokes shift and liquid-like dielectric solvation of band edge carriers in lead-halide perovskites. *Nat. Commun.* **10,** 1–8 (2019).

85. Guo, P. *et al.* Infrared-pump electronic-probe of methylammonium lead iodide reveals electronically decoupled organic and inorganic sublattices. *Nat. Commun.* **10,** (2019).

86. Zheng, F. & Wang, L. W. Large polaron formation and its effect on electron transport in hybrid perovskites. *Energy Environ. Sci.* **12,** 1219–1230 (2019).

87. Dalpian, G. M., Zhao, X. G., Kazmerski, L. & Zunger, A. Formation and Composition-Dependent Properties of Alloys of Cubic Halide Perovskites. *Chem. Mater.* **31,** 2497–2506 (2019).

88. Wu, B. *et al.* Indirect tail states formation by thermal-induced polar fluctuations in halide perovskites. *Nat. Commun.* **10,** 1–10 (2019).

89. Joshi, P. P., Maehrlein, S. F. & Zhu, X. Dynamic Screening and Slow Cooling of Hot Carriers in Lead Halide Perovskites. *Adv. Mater.* **1803054,** 1–10 (2019).

90. Lan, Y. *et al.* Ultrafast correlated charge and lattice motion in a hybrid metal halide perovskite. *Sci. Adv.* **5,** 5558 (2019).

91. Yin, J. *et al.* Point Defects and Green Emission in Zero-Dimensional Perovskites. *J. Phys. Chem. Lett.* **9,** 5490–5495 (2018).

92. Perdew, J. P., Burke, K. & Ernzerhof, M. Generalized gradient approximation made simple. *Phys. Rev. Lett.* **77,** 3865–3868 (1996).

93. Tkatchenko, A. & Scheffler, M. Accurate molecular van der Waals interactions from ground-state electron density and free-atom reference data. *Phys. Rev. Lett.* **102,** 073005 (2009).

94. Bučko, T., Lebègue, S., Hafner, J. & Ángyán, J. G. Improved density dependent correction for the description of London dispersion forces. *J. Chem. Theory Comput.* **9,** 4293–4299 (2013).

95. Bučko, T., Lebègue, S., Ángyán, J. G. & Hafner, J. Extending the applicability of the Tkatchenko-Scheffler dispersion correction via iterative Hirshfeld partitioning. *J. Chem. Phys.* **141,** 0–17 (2014).

96. Heyd, J., Scuseria, G. E. & Ernzerhof, M. Hybrid functionals based on a screened Coulomb potential. *J. Chem. Phys.* **118,** 8207–8215 (2003).

97. Kresse, G. & Furthmüller, J. Efficient iterative schemes for ab initio total-energy calculations using a plane-wave basis set. *Phys. Rev. B - Condens. Matter Mater. Phys.* **54,** 11169–11186 (1996).

98. Joubert, D. From ultrasoft pseudopotentials to the projector augmented-wave method. *Phys. Rev. B - Condens. Matter Mater. Phys.* **59,** 1758–1775 (1999).

99. Egger, D. A. & Kronik, L. Role of dispersive interactions in determining structural properties of organic-inorganic halide perovskites: Insights from first-principles calculations. *J. Phys. Chem. Lett.* **5,** 2728–2733 (2014).

100. Wang, Y. *et al.* Density functional theory analysis of structural and electronic properties of orthorhombic perovskite $CH_3NH_3PbI_3$. *Phys. Chem. Chem. Phys.*

**16,** 1424–1429 (2014).

101. Li, J. & Rinke, P. Atomic structure of metal-halide perovskites from first principles: The chicken-and-egg paradox of the organic-inorganic interaction. *Phys. Rev. B* **045201,** 12 (2016).

102. Beck, H., Gehrmann, C. & Egger, D. A. Structure and binding in halide perovskites: Analysis of static and dynamic effects from dispersion-corrected density functional theory. *APL Mater.* **7,** (2019).

103. Stoumpos, C. C. *et al.* Crystal growth of the perovskite semiconductor CsPbBr$_3$: A new material for high-energy radiation detection. *Cryst. Growth Des.* **13,** 2722–2727 (2013).

104. Nosé, S. A unified formulation of the constant temperature molecular dynamics methods. *J. Chem. Phys.* **81,** 511–519 (1984).

105. Hoover, W. G. Constant-pressure equations of motion. *Phys. Rev. A* **34,** 2499–2500 (1986).

106. Jones, R. O. & Gunnarsson, O. The density functional formalism, its applications and prospects. *Rev. Mod. Phys.* **61,** 689–746 (1989).

107. Mladenović, M. & Vukmirović, N. Effects of thermal disorder on the electronic structure of halide perovskites: insights from MD simulations. *Phys. Chem. Chem. Phys.* (2018).

108. Mayers, M., Tan, L. Z., Egger, D. A., Rappe, A. M. & Reichman, D. R. How Lattice and Charge Fluctuations Control Carrier Dynamics in Halide Perovskites. *Nano Lett.* **18,** 8041–8046 (2018).

109. Nan, G. *et al.* How Methylammonium Cations and Chlorine Dopants Heal Defects in Lead Iodide Perovskites. *Adv. Energy Mater.* **8,** 1702754 (2018).

110. Even, J., Pedesseau, L., Jancu, J. M. & Katan, C. Importance of Spin-Orbit Coupling in Hybrid Organic / Inorganic Perovskites for Photovoltaic Applications. *J. Phys. Chem. Lett.* **4,** 2999−3005 (2013).

111. Lim, A. *et al.* Electron Elevator: Excitations across the Band Gap via a Dynamical Gap State. *Phys. Rev. Lett.* **116,** (2016).

112. Krause, D. & Thörnig, P. JURECA: General-purpose supercomputer at Jülich Supercomputing Centre. *J. large-scale Res. Facil. JLSRF* **4,** A132 (2018).

**Figures:**

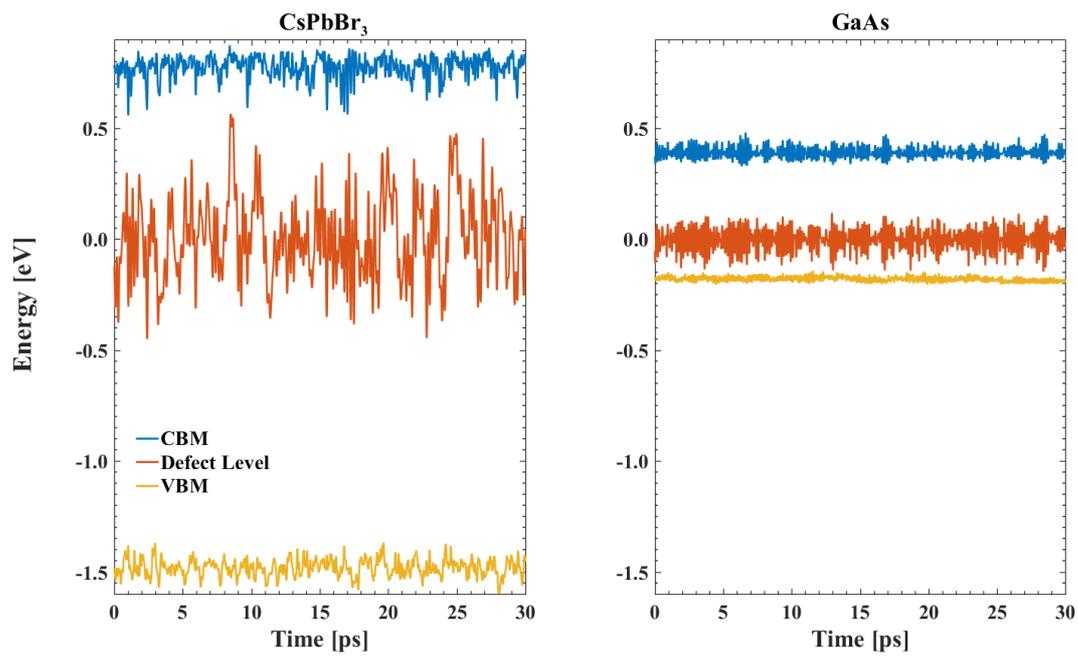

Figure 1: Eigenvalues representing the valence band maximum (VBM), the defect level, and the conduction band minimum (CBM) as a function of time along the MD trajectory, for (a) CsPbBr$_3$ with V$_{Br}$ and (b) GaAs with As$_{Ga}$. All values are referenced to the average defect level position. For clarity, only the last 30 ps of the MD run of each material are shown.

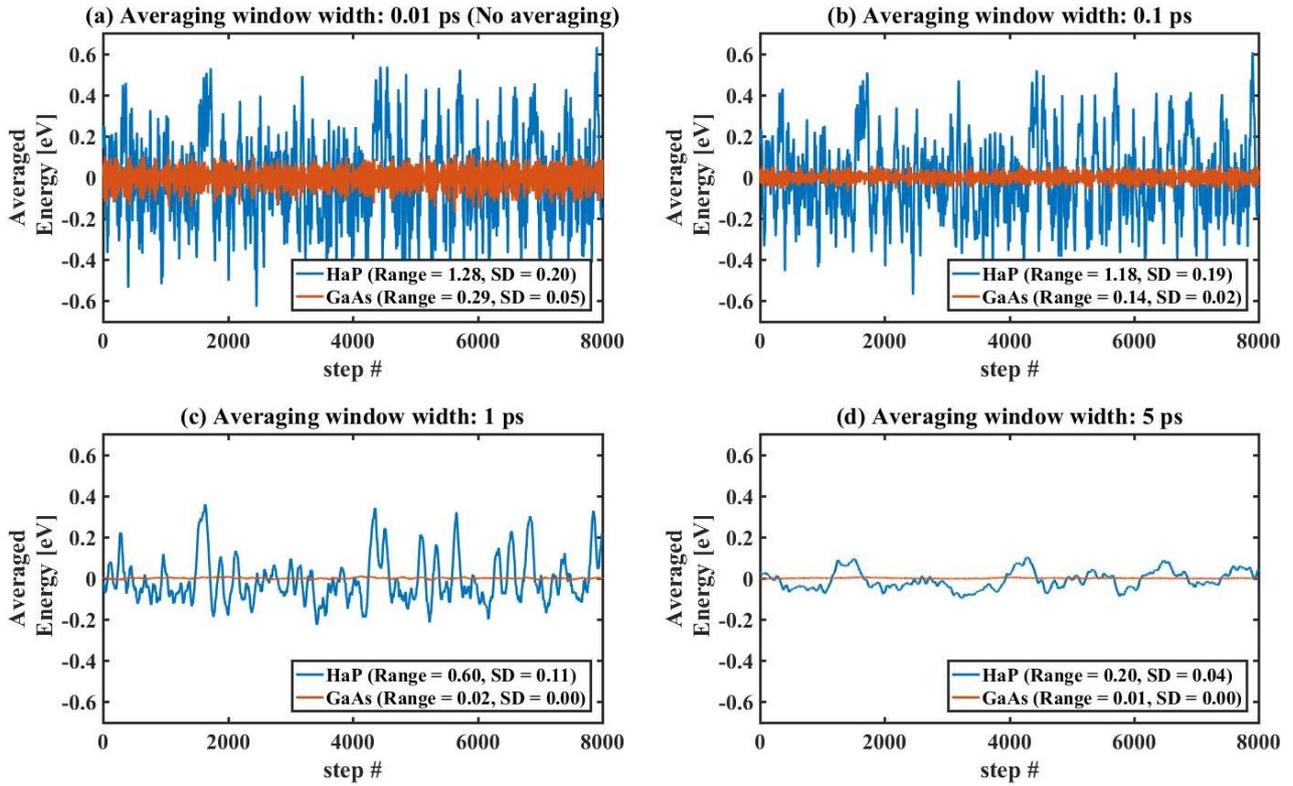

Figure 2: Averaged defect level energy of the CsPbBr$_3$ and GaAs supercells along the MD trajectory, (a) as obtained from the simulation, and (b-d) obtained by using a moving averaging temporal window of width 0.1 ps, 1 ps, and 5 ps. The range of fluctuations and standard deviation (SD) for each averaging window are given in the legend. The ratio of defect level ranges in the two materials grows from 4:1 up to > 20:1 as the window width meets and exceeds 1 ps, showing that the CsPbBr$_3$ defect energy level has dramatically greater energy scatter over experimentally-accessible time scales.

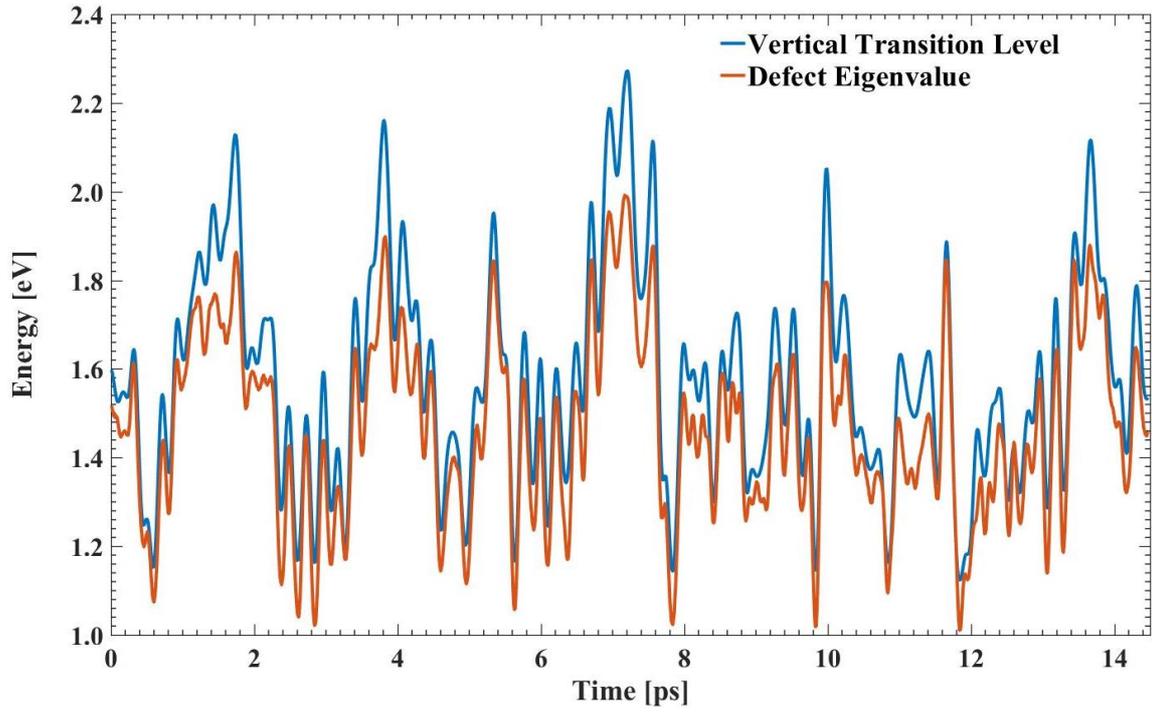

Figure 3: Eigenvalues representing the neutral defect level (orange line), referenced to the valence band maximum (VBM), compared to the vertical transition level of a Br vacancy (taken as the total energy difference between a supercell with a neutral defect and a positively charged one, blue line), in a Br-vacancy containing 2x2x2 CsPbBr$_3$ supercell, as a function of time along the MD trajectory for 15 ps of the MD simulation.

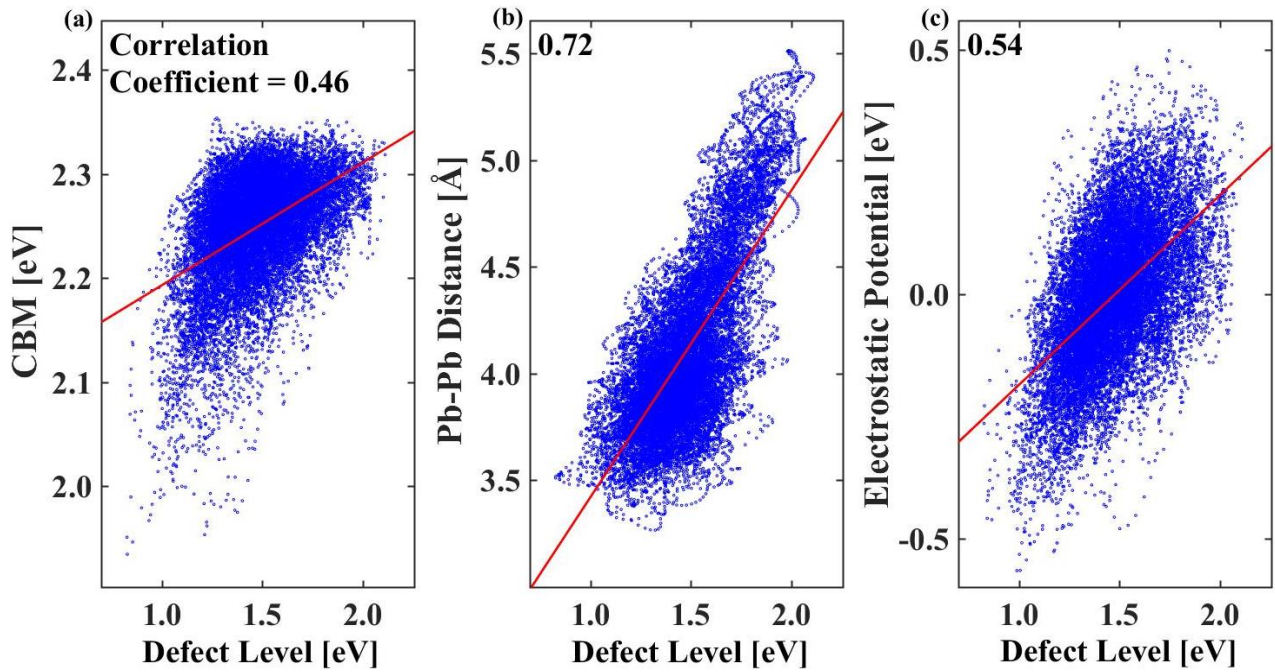

Figure 4: Correlation between the defect level fluctuations along the MD trajectory and (a) CBM, (b) distance between the two Pb atoms that surround the Br vacancy, and (c) electrostatic potential at these Pb sites. The correlation coefficient for each two variables is given at the upper left corner of each panel. The electrostatic potential is referenced to its own average value, while the CBM and the defect level are referenced to the average VBM.

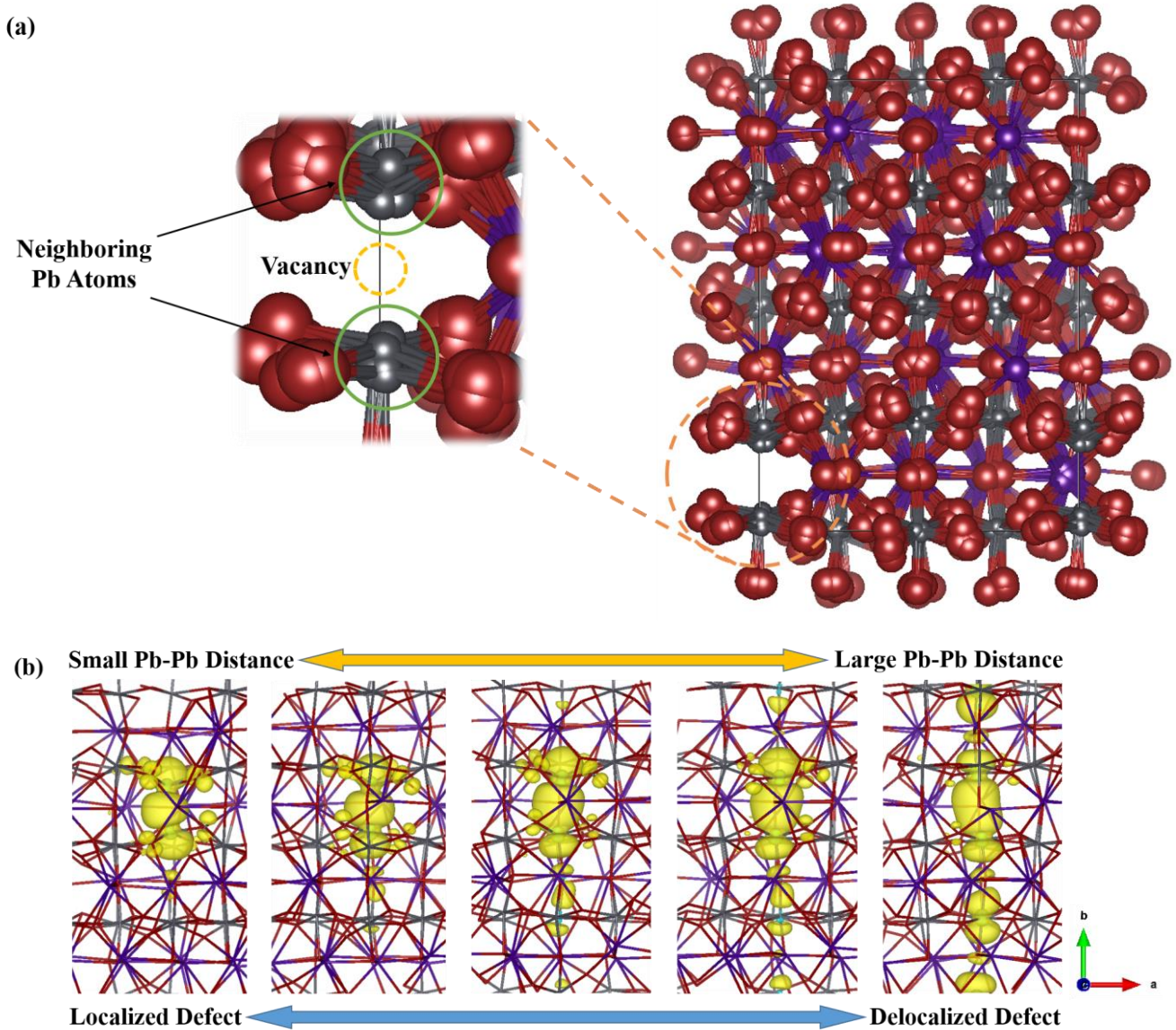

Figure 5: (a) A representation of the dynamic nature of the CsPbBr$_3$ system, shown by overlaying five different geometries from the MD trajectory. Cs atoms are in purple, Pb in grey, and Br in red. The location of the vacancy is marked by a yellow dashed circle, and its neighboring Pb atoms are marked by a green circle. (b) The partial charge density of the eigenvalue associated with the defect level (yellow surface), calculated for the same five geometries. The distance between the two Pb atoms that surround the Br vacancy is smallest for the left-most geometry and largest for the right-most geometry. The opposite is true for the amount of charge localization.

**Tables:**

Table 1: Statistical analysis of the eigenvalues representing the valence band maximum (VBM), defect level ($E_d$), conduction band minimum (CBM), and CBM-$E_d$, performed on the results of the entire 180 ps MD run of the $V_{Br}$-containing $CsPbBr_3$ supercell.

|  | VBM [eV] | $E_d$ [eV] | CBM [eV] | CBM – $E_d$ [eV] |
|---|---|---|---|---|
| **Minimum** | 0.46 | 1.43 | 2.53 | 0.17 |
| **Maximum** | 0.76 | 2.71 | 2.95 | 1.29 |
| **Range** | 0.29 | 1.28 | 0.42 | 1.12 |
| **Average** | 0.60 | 2.08 | 2.85 | 0.77 |
| **Standard Deviation** | 0.04 | 0.20 | 0.05 | 0.18 |